\documentclass{ws-procs10x7}
\usepackage{balance}

\newcolumntype{d}[1]{D{.}{.}{#1}}

\makeindex
\begin{document}
\def\KS{K^0_S}
\def\Xipi{\Xi(1321)\pi}
\def\Dstarp{D^{\ast}p(3100)}
\def\dstar{D^{\ast}}
\hyphenation{re-so-nance}
\hyphenation{doub-ly}

\title{PENTAQUARK \ SEARCHES \  IN \  H1}

\author{J.E. Olsson$^*$}

\address{DESY, Hamburg, Germany \\$^*$E-mail: jan.olsson@desy.de \\ 
Talk given on behalf of the H1 Collaboration at ICHEP06, Moscow, Russia}

\twocolumn[\maketitle\abstract{We report on searches in deep inelastic 
$ep$ scattering for narrow baryonic 
states decaying into $\Xi^-\pi^-, \Xi^-\pi^+, \KS p$ and their charge 
conjugates, at centre-of-mass energies of 300 and 318 GeV. No signal for a
new narrow 
baryonic state is observed in the mass ranges 1600-2100 MeV ($\Xi\pi$) and
from threshold up to 1700 MeV ($\KS p$). The standard baryon 
$\Xi(1530)^0$ is observed in the decay mode $\Xi^-\pi^+$, and mass
dependent upper 
limits on the ratio of the hypothetical pentaquark states $\Xi^{--}_{5q}$
and  $\Xi^0_{5q}$ to the $\Xi(1530)^0$ signal are given. Also for
the hypothetical strange pentaquark $\Theta^+$ mass 
dependent upper limits on $\sigma(ep\rightarrow e\Theta^+X)\times
 BR(\Theta^+\rightarrow K^0p)$ are obtained. 
\newline
Finally measurements of the acceptance corrected ratios 
$\sigma(D^{\ast}p(3100))/\sigma(D^{\ast})$ for the electroproduction of the
anti-charmed baryon state $D^{\ast}p(3100)$ decaying into $D^{\ast}$ and $p$ 
are presented.
}
]

\section{Introduction}

Pentaquarks are exotic to the standard quark model of mesons and baryons,
although not excluded in QCD. Combination of the meson and baryon
octets leads to the anti-decuplet $qqqq\bar{q}$, where $q$ stands for the 
light quarks $u,d$ and $s$.
These 5-quark states are non-minimal, colour neutral combinations.
The apex states $uudd\bar{s},  ssdd\bar{u}$ and $uuss\bar{d}$ are 
predicted to have masses in the range $1.5 - 2.1$ GeV and 
to be very narrow\cite{Diakonov}. 
For two of 
these states decay modes are experimentally relatively easily accessible, 
and consequently many searches have been performed, both in fixed target
and colliding beam environments. For the  $uudd\bar{s}$ state, called
$\Theta^+$, a large number of observations in the mass range $1.52-1.54$~GeV
is matched by an equally large number of non-observations\cite{Hicks}.
The doubly strange $ssdd\bar{u}$ state, $\Sigma^{--}$, has only been
observed by one experiment so far\cite{NA49}. Also here many
non-observations have been reported\cite{Hicks}. 
\par
Pentaquarks which contain a charm quark have also been searched for. One
observation\cite{H1Dstarp}, named $D^{\ast}p(3100)$ by the H1 collaboration, 
has so far not been confirmed by any other experiment.
\par
In this report the current results of searches 
by the H1 experiment for the $\Theta^+$ and
$\Sigma^{--}$ states are briefly presented. 
The searches have been performed in the
HERA-I $ep$ data (1996-2000), which encompass $75-100$ pb$^{-1}$.
In the last section some details of the characteristics of 
the $D^{\ast}p(3100)$  production are presented.    

\section{The strange pentaquark $\Theta^+$}

The $\Theta^+$ was first seen by the LEPS collaboration\cite{LEPS} in the decay
mode $K^+n$, at a mass of 1.52 GeV. 
The H1 collaboration searched\cite{H1Thetapubl} for a narrow state
decaying to $\KS p$ in the
mass interval from threshold 1.48 GeV to 1.7 GeV, using DIS data with
$5<Q^2<100$ GeV$^2$ and $0.1<y<0.6$. The $\KS$
was identified through the decay $\KS\rightarrow\pi^+\pi^-$
and events were accepted if they contained at least one $\KS$  
and at least one proton\footnote{Charge conjugation is always implied, unless
otherwise stated.} 
candidate. 
Charged tracks had $p_t>0.15$ GeV and pseudorapidity  $|\eta|<1.75$, 
while the $K_S^{\circ}$ candidate had $p_t>0.3$ GeV. Backgrounds from 
$\Lambda$
and converted photons were rejected with the restrictions 
$M_{p\pi}>1.125$ GeV and $M_{ee}>0.05$ GeV on the $\pi\pi$ system. 
Protons were identified through the specific ionization 
loss $dE/dx$ in the inner drift chambers, with efficiencies between 
65 and 100\%.
\smallskip
\par\noindent
\begin{minipage}[t]{7cm}
\begin{minipage}[t]{6.9cm}
\epsfxsize=3.5cm\epsfbox{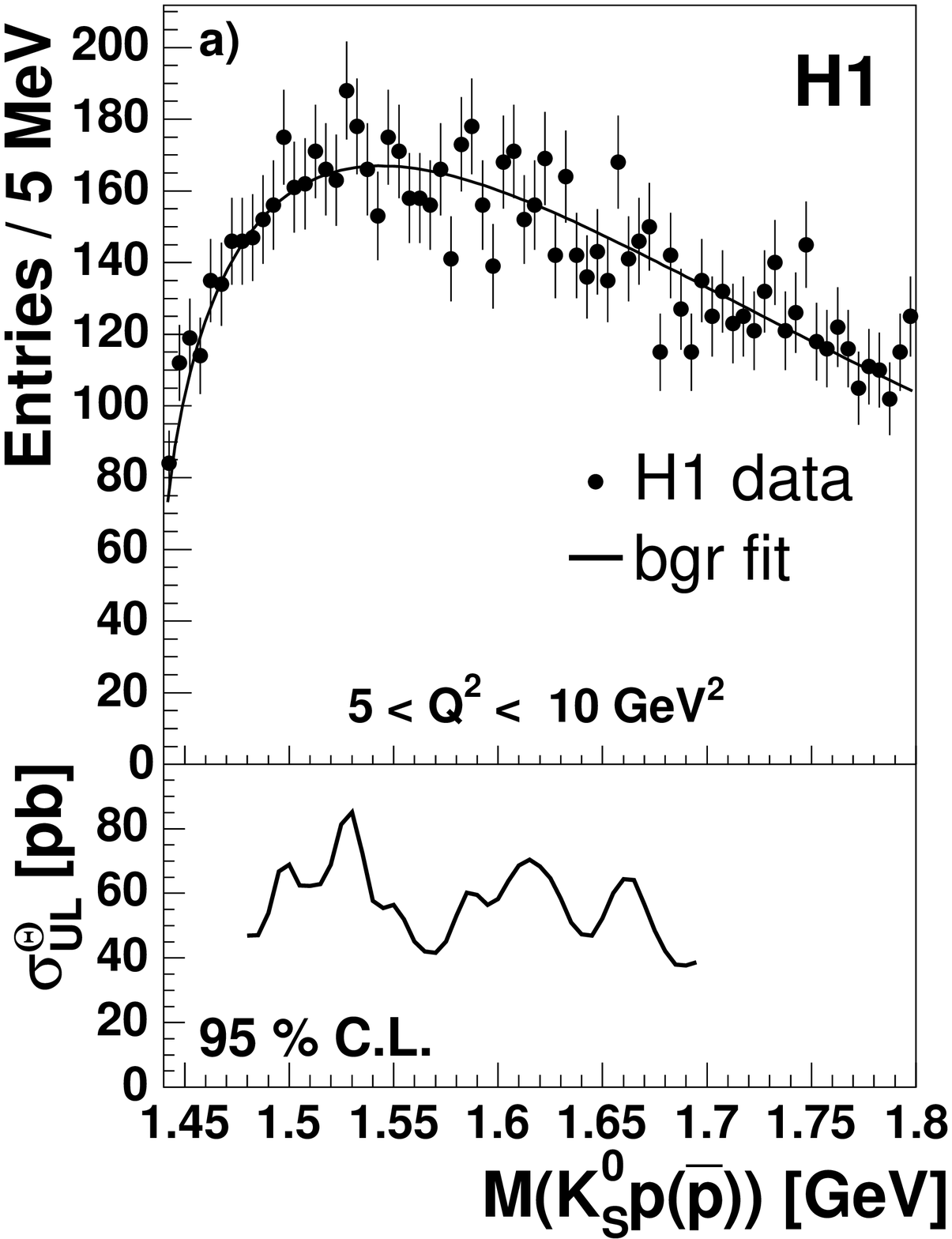}
\vspace*{-4.55cm}
\par\noindent
\hspace*{3.4cm}
\epsfxsize=3.5cm\epsfbox{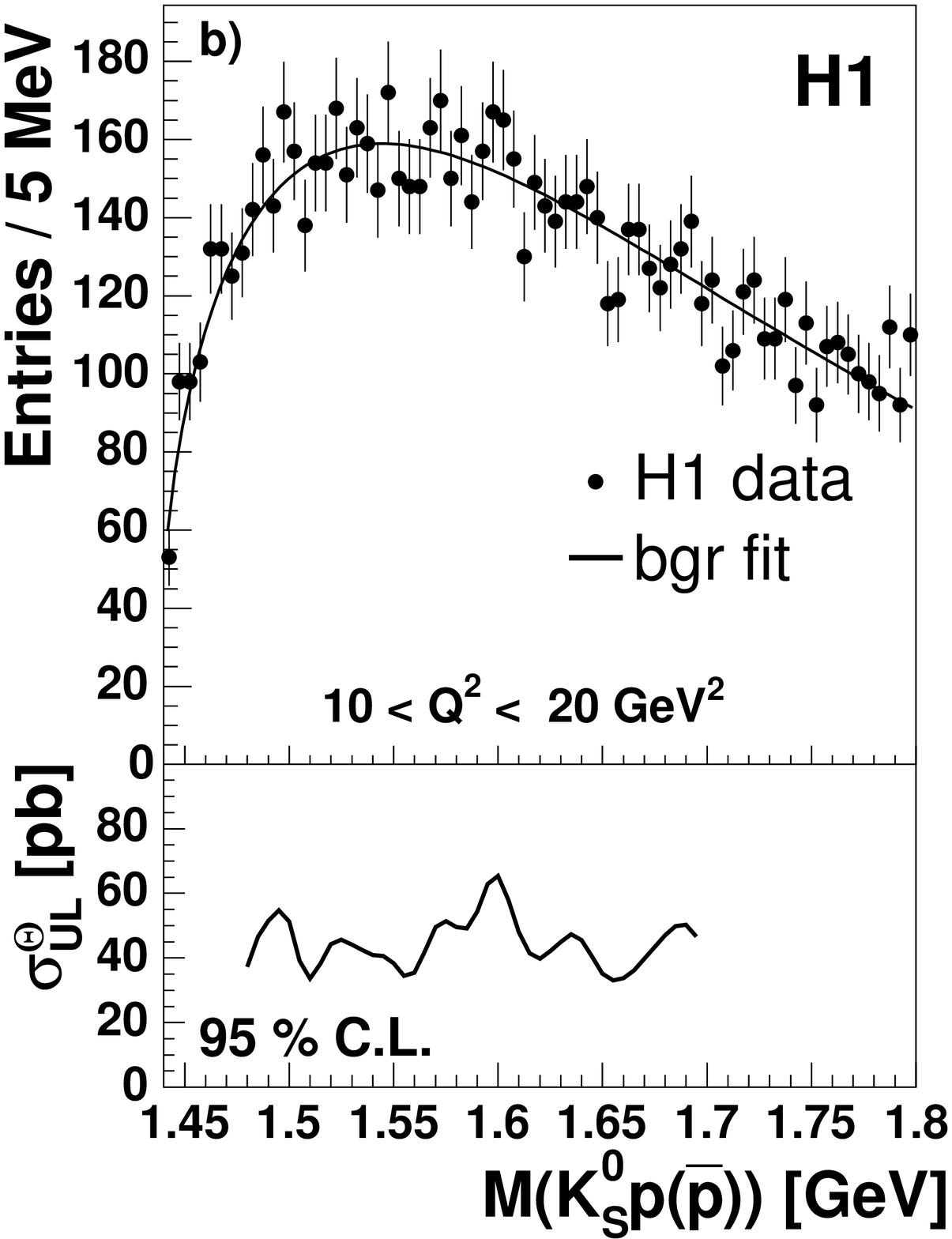}
\par\noindent
\epsfxsize=3.5cm\epsfbox{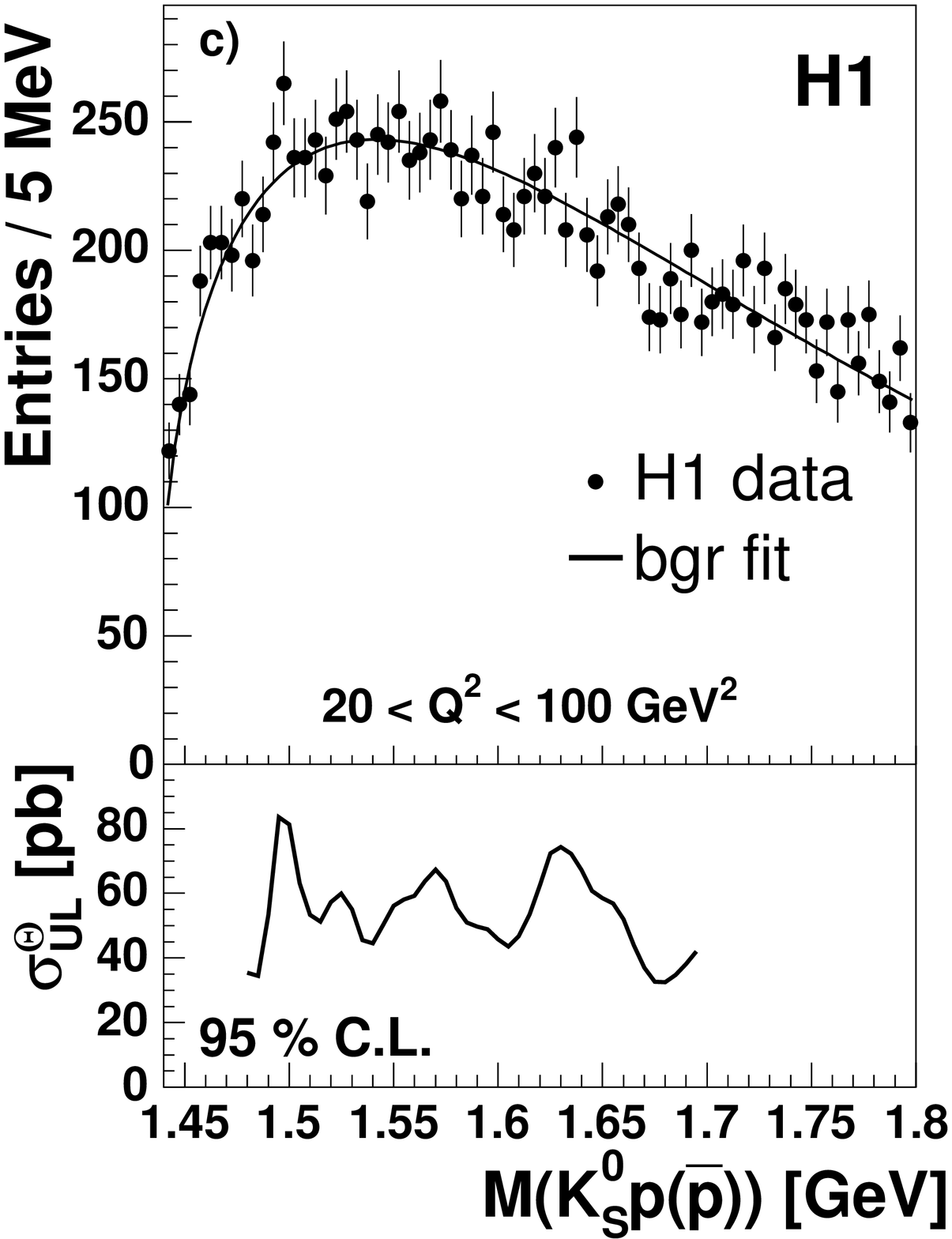}
\vspace*{-4.55cm}
\par\noindent
\hspace*{3.4cm}
\epsfxsize=3.5cm\epsfbox{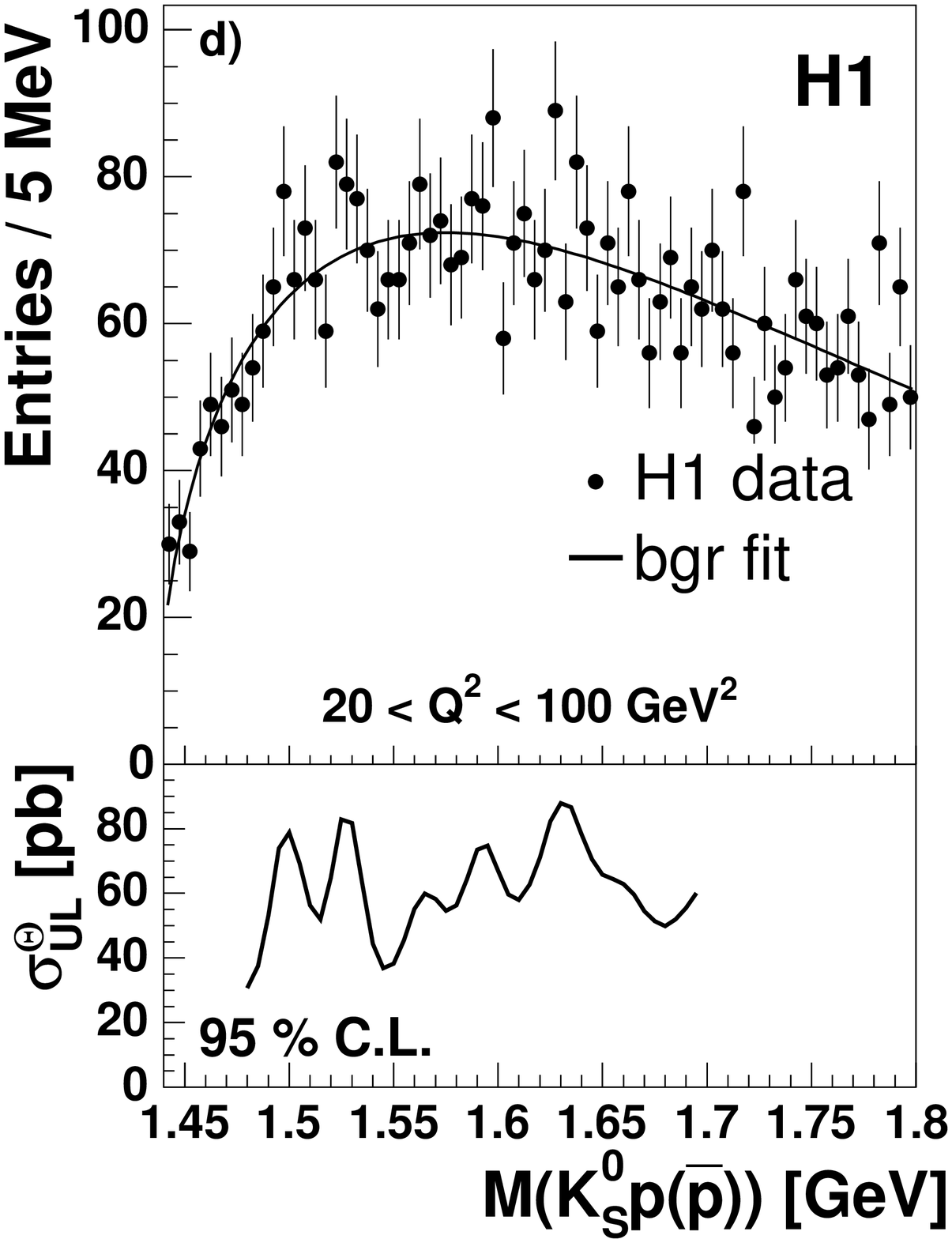}
\end{minipage}
\smallskip
\par\noindent
\hspace*{0.4cm}
\begin{minipage}[t]{6cm}
 {Fig. 1. \ \ $K^0_Sp(\bar{p})$ invariant mass. a-c) in bins of $Q^2$, d) in
the highest $Q^2$ bin and with $p_p<1.5$~GeV. The full curves show the 
background function fits. Upper limits $\sigma_{U.L.}$ at 95\% C.L. on the 
cross section are shown below the mass distributions.
}
\end{minipage}
\end{minipage}
\smallskip
\par
In the resonance search, $\KS p$ combinations were selected under 
the restriction
$p_t>0.5$ GeV and $|\eta|<1.5$ for the  $\KS p$ system. 
Mass distributions $\KS p$ in three
$Q^2$ intervals are shown in Fig.~1a-c. 
No significant resonance signal is seen, and
the distributions are well described by a smooth background function. 
Mass dependent upper limits for the cross section 
$\sigma(ep\rightarrow e\Theta X)\times BR(\Theta\rightarrow K^0p)$,
obtained using
a modified frequentist approach\cite{NIMA434} based on likelihood ratios
and taking into account both statistical and systematic 
errors,
are shown below the mass distributions in Fig.~1.
\par
The search was repeated with separation of the $\KS p$ and 
$\KS\bar{p}$ 
distributions, with no significant peak as result. 
The obtained upper limits vary between 30 and 90~pb.
\par
The ZEUS collaboration has reported evidence\cite{ZEUSTheta} for a 
1.52 GeV signal in the $\KS p$ mass distribution. The H1 upper limit
at 1.52 GeV is $\sigma~<~72$~pb, which translates to 
$\sigma~<~100$~pb (95\% C.L.) when extrapolated to the ZEUS $y$-range.
This value is barely compatible with the
ZEUS preliminary cross section\cite{ZEUSThetaprel},
$\sigma = 125\pm 27^{+36}_{-28}$ pb.
\par
The ZEUS collaboration also found that the observed 
resonance is most prominent with the cuts $Q^2>20$~GeV and  
proton momentum $p_p <1.5$ GeV. Also in this kinematic range 
the H1 collaboration observes no peak, see Fig.~1d.

\section{The pentaquarks $\Xi^{--}_{5q}$
                               and  $\Xi^0_{5q}$}

The NA49 collaboration 
observed a narrow resonance structure at $1862\pm 2$ MeV 
in $\Xipi$ mass spectra\cite{NA49}. 
Both $\Xi^{--}$ and $\Xi^0$ peaks were seen, 
leading to the interpretation of these states being the neutral and 
doubly charged members of the $\Xi_{3/2}(1862)$ pentaquark multiplet. 
Other experiments could not yet confirm this observation.
\par
The H1 search for this state\cite{H1Xiprel} uses the decay chain 
$\Xi \rightarrow \Xipi, \ \Xi(1321)\rightarrow \Lambda\pi, \ 
\Lambda\rightarrow p\pi$. 
DIS events were selected with $2<Q^2<120$ GeV$^2$ and $0.05<y<0.7$.
In a mass window of $\pm~8$~MeV
$\sim~158000~\Lambda$ candidates were identified with a 3-dimensional 
vertex fit, using cuts on  the momentum $p_{t,p\pi}>0.3$ GeV and 
decay length $>0.75$ cm.  
$\Lambda\pi$ combinations were also subjected to a 3-dimensional vertex fit, 
with a
further cut on the distance of closest approach
to the primary vertex.
The significance of the $\Xi (1321)$ signal is increased by a restriction to
$<0.6$ rad on the angle
between secondary and tertiary vertex vectors. In the resulting 
mass distribution  of $\Lambda\pi, \sim~1650~\Xi(1321)$ form a 
clear narrow peak.
\par 
Finally the $\Xi(1321)$, in a mass window of $\pm~15$ MeV, 
was combined with a charged pion
from the primary vertex. A cut $p_t>1.0$ GeV was 
imposed on the $\Xipi$ combinations.
Fig. 2 shows the final mass distributions.
\par
In the neutral $\Xipi$ combinations there is a clear signal of the wellknown
$\Xi(1530)^0$, with $\sim~170$ events. 
The doubly charged $\Xipi$ combinations do not show
any resonant structure, in particular not at 1.86 GeV, the mass of the NA49
observation. This is also true when separating $\Xi^-\pi^-$ and 
$\Xi^+\pi^+$ distributions.
The $\Xi(1530)^0$ is well seen in both neutral charge combinations.
\smallskip
\par\noindent
\begin{minipage}[t]{7cm}
\begin{minipage}[t]{6.9cm}
\epsfxsize=3.5cm\epsfbox{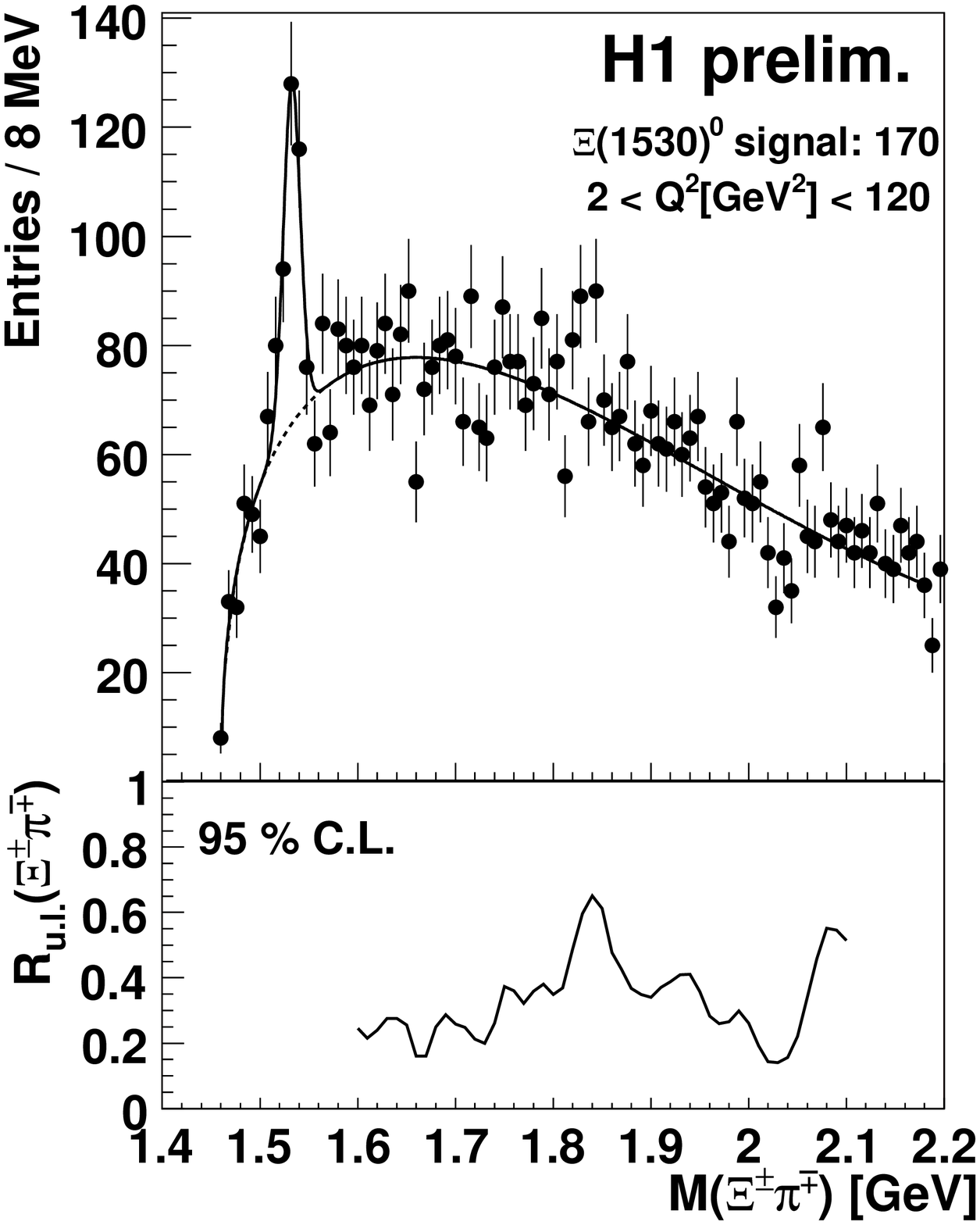}
\vspace*{-4.4cm}
\par\noindent
\hspace*{3.4cm}
\epsfxsize=3.5cm\epsfbox{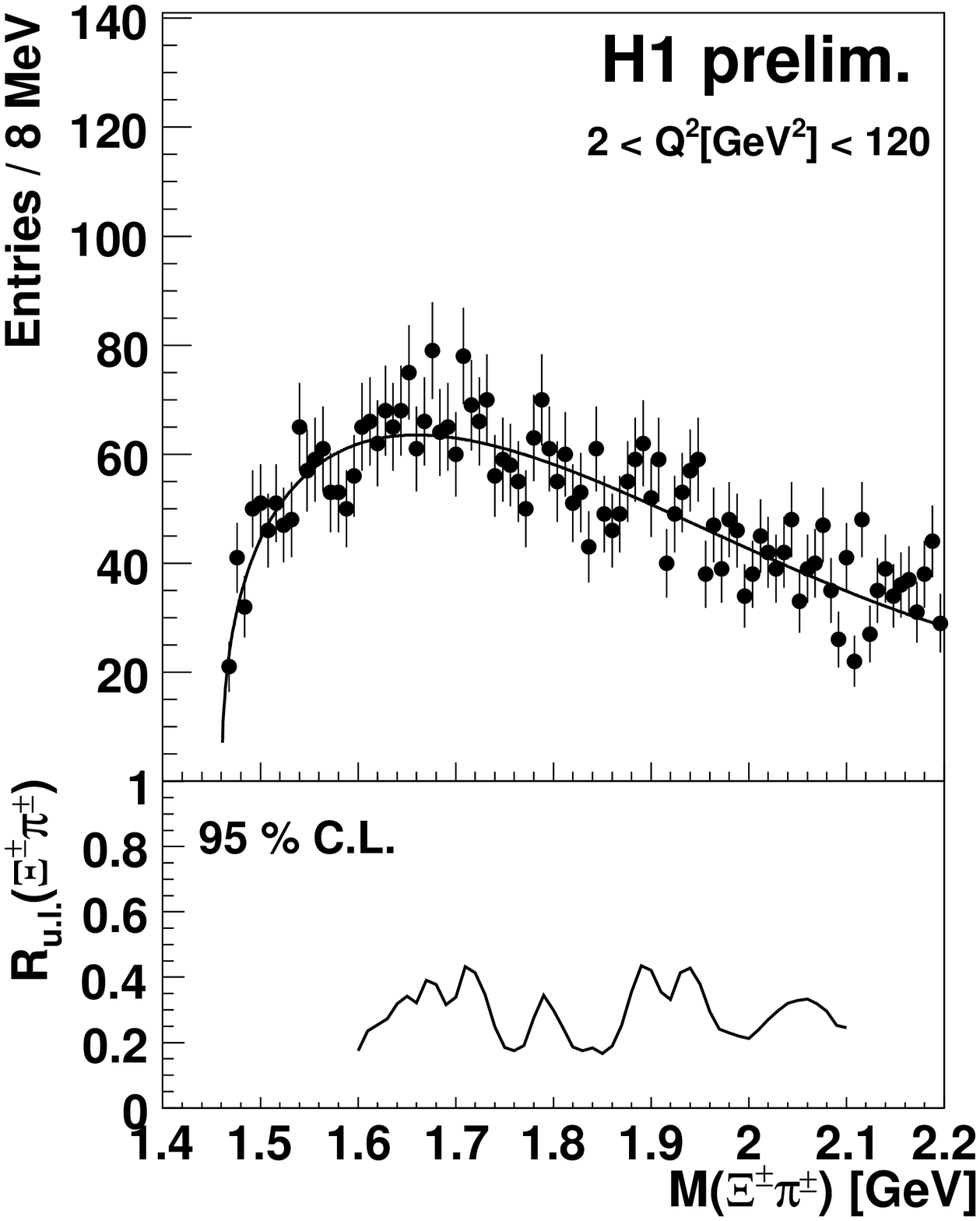}
\end{minipage}
\vspace*{-4.0cm}
\par\noindent
\hspace*{1.2cm} {\bf a)} \hspace*{2.3cm} {\bf b)}
\vspace*{3.6cm}
\par\noindent
\hspace*{0.4cm}
\begin{minipage}[t]{6cm}
 {Fig. 2. \ \ $\Xi\pi$ invariant mass, summed for a) two opposite and b) two
equal charge combinations. Solid lines show fits of a background function,
in a) including a Gaussian. Upper limits $R_{U.L.}$ at 95\% C.L. on the 
ratio of the number of events of a hypothetical $\Xipi$ resonance 
to the $\Xi(1530)^0$ are shown below the mass distributions. 
}
\end{minipage}
\end{minipage}
\smallskip
\par
Mass dependent upper limits are defined in terms of the 
ratio of a hypothetical $\Xipi$ resonance to the $\Xi(1530)^0$, 
using a narrow Gaussian 
for a possible signal in the range $1.6-2.1$ GeV. The background is a smooth
function and again the modified frequentist approach\cite{NIMA434} is
used. Separate upper limits were obtained for neutral and doubly charged 
combinations, as well as for their sum. The ratio limits are shown in the lower
plots of Fig.~2 and lie in the range $0.15 - 0.6$, with the value 
$R_{U.L.}(1860) \sim~0.5 (0.2)$ for the neutral (doubly charged) combination. 
Summing all combinations,
the limit $R_{U.L.}(1860) \sim~0.5$ is obtained, which is fully compatible 
with the upper limit value 0.29, obtained by the ZEUS experiment in a similar 
analysis\cite{ZEUSXi}.

\section{$D^{\ast}p(3100)$~production in DIS}

Pentaquark multiplets containing the heavier $c$ or $b$ quarks have also been
considered\cite{refThetac}. 
If the anti-charmed pentaquark $\Theta_c^0$, with the quark
content $uudd\bar{c}$, is heavy enough the
decay $\Theta_c^0 \rightarrow D^{\ast -} p$ would be possible. 
Evidence for a narrow peak, provisionally labelled $\Dstarp$,
in the $\dstar p$ mass distribution in DIS and photoproduction has been given 
by the H1 collaboration\cite{H1Dstarp}. 
In subsequent searches, no other experiment was able 
to confirm this  observation. 

Additional preliminary information for the $\Dstarp$ production
is provided by the 
H1 collaboration\cite{H1Dstarpprel}.  
Acceptance corrected yield ratios relative to inclusive $\dstar$ production,
and differential distributions of the visible cross section ratio as function
of event kinematics and $\dstar$ quantities are presented, the latter hinting
at some features of the $\Dstarp$ production mechanism.
\smallskip
\par
The acceptance corrected yields ratio $R_{cor}(\Dstarp/\dstar)$ 
is defined in the visible
range given by $p_t(\Dstarp)>1.5$~GeV, $-1.5<\eta(\Dstarp)<1.0$,  
$p_t(\dstar)>1.5$~GeV, $-1.5<\eta(\dstar)<1.0$ and $z(\dstar)>0.2$ 
(including the $\dstar$ from the $\Dstarp$ decay). $\eta$ and $z$ are the
pseudorapidity and elasticity, respectively. The 
acceptance corrections are calculated using RAPGAP, 
under the assumption that
pentaquarks are produced by the fragmentation (simulated with the Lund string
model). 
The observed yields ratio $R(\Dstarp/\dstar) = 1.46\pm 0.32$\% becomes, 
after the acceptance correction, 
$R_{cor}(\Dstarp/\dstar) = 1.59\pm 0.33^{+0.33}_{-0.45}$.
The ZEUS 
collaboration found $R_{cor} < 0.59$~\% (at 95\% C.L.),
using larger statistics and a different definition of the 
visible range\cite{ZEUSDstarp}.
\smallskip 
\par\noindent
\begin{minipage}[t]{7cm}
\begin{minipage}[t]{6.9cm}
\epsfxsize=3.5cm\epsfbox{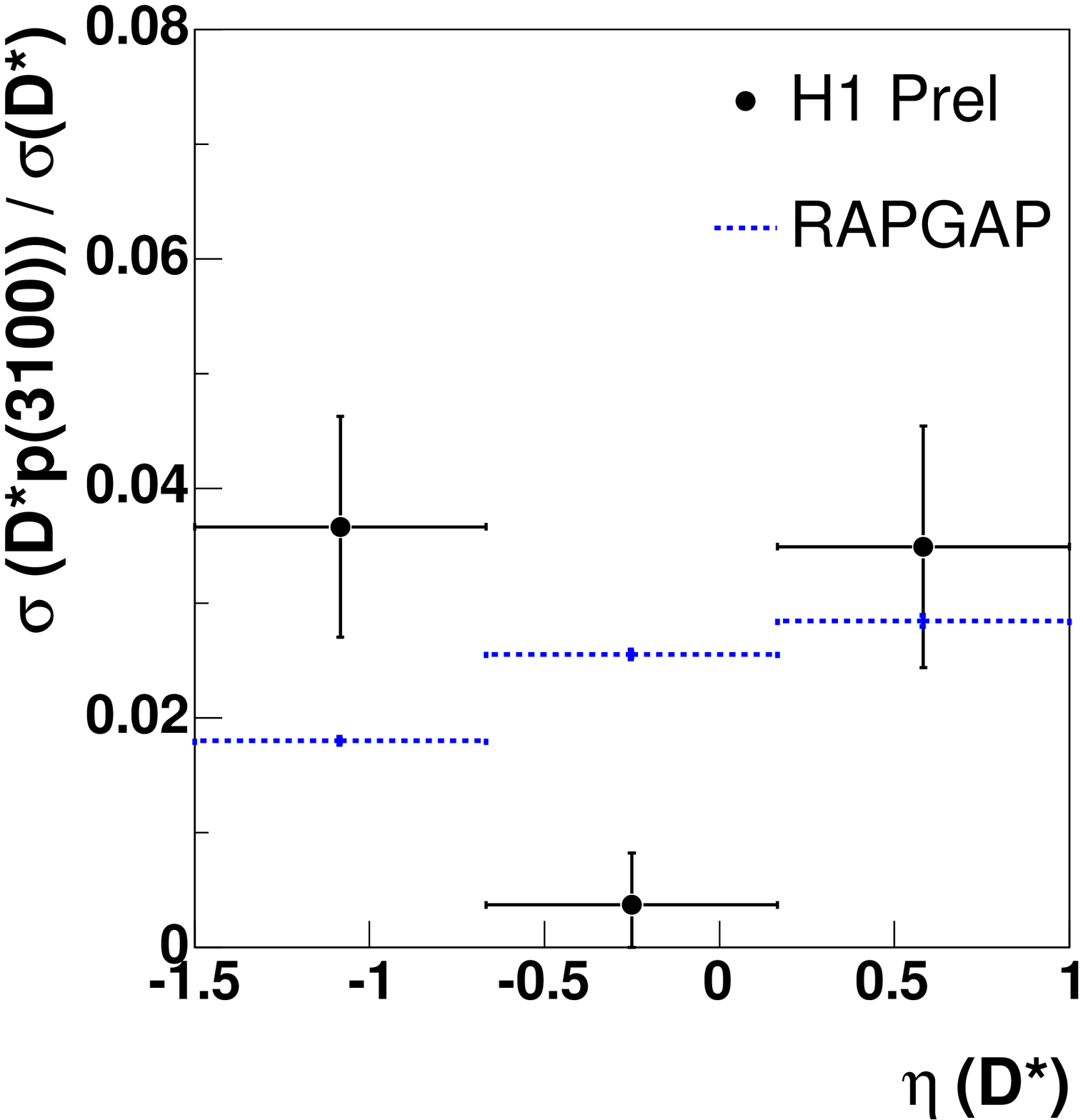}
\vspace*{-3.55cm}
\par\noindent
\hspace*{3.4cm}
\epsfxsize=3.5cm\epsfbox{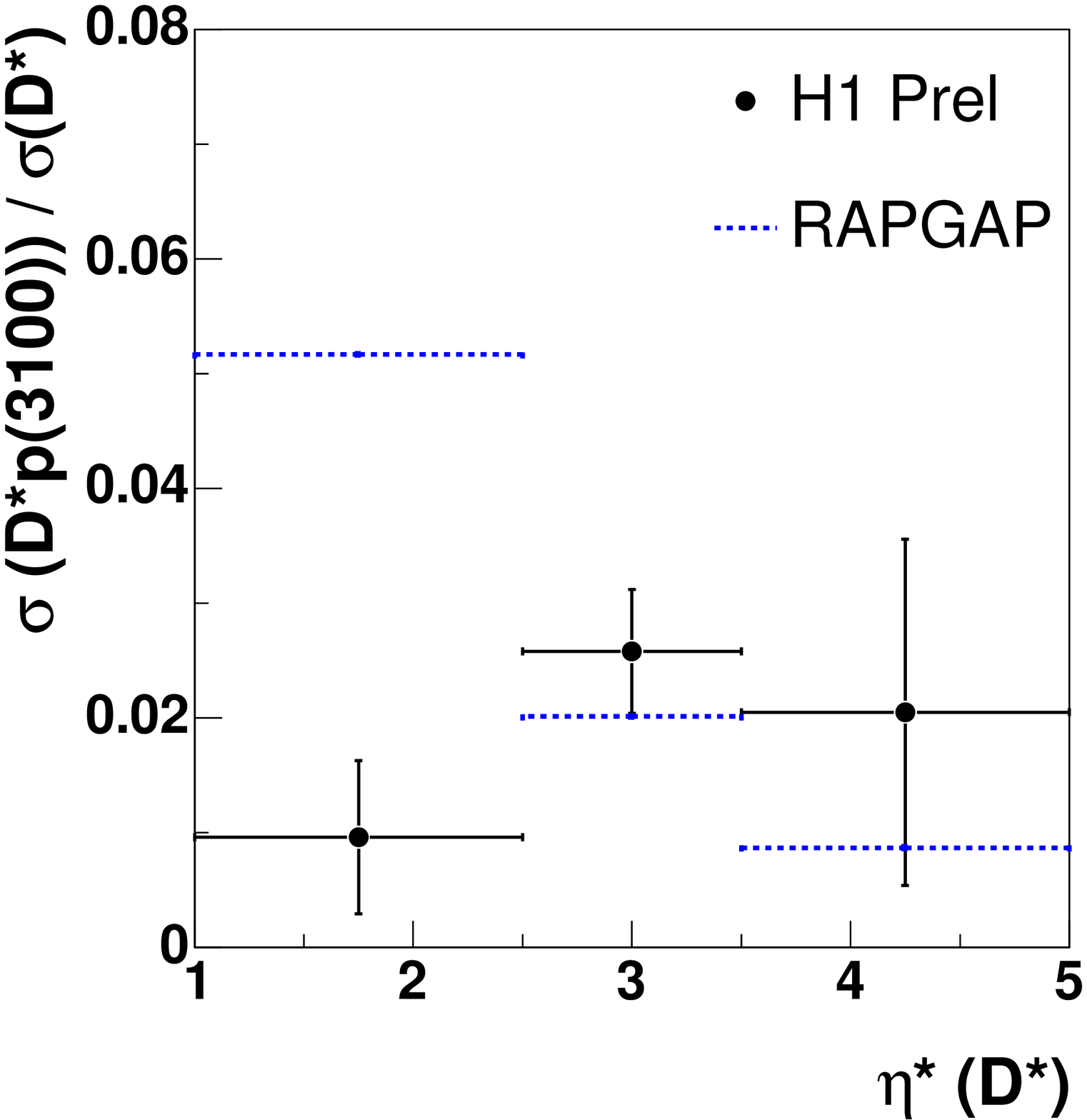}
\end{minipage}
\vspace*{-3.3cm}
\par\noindent
\hspace*{0.7cm} {\bf a)} \hspace*{3cm} {\bf b)}
\vspace*{2.9cm}
\par\noindent
\hspace*{0.4cm}
\begin{minipage}[t]{6cm}
 {Fig. 3. \ \ Acceptance corrected ratio 
$\sigma_{vis}(\Dstarp)/\sigma_{vis}(\dstar)$ as a function of the 
$\dstar$ pseudorapidity, in the a) laboratory and b) hadronic c.m. system.
Only statistical errors are shown.
}
\end{minipage}
\end{minipage}
\smallskip
\par
Extrapolating to the full phase space, the visible 
cross section ratio is obtained: 
$\sigma_{vis}(\Dstarp)/\sigma_{vis}(\dstar) = 2.48\pm 0.52^{+0.85}_{-0.64}$~\%.
The differential distribution of this cross section ratio 
in the variables $\eta(\dstar)$ and  $\eta^{\ast}(\dstar)$ (i.e. 
pseudorapidity in the laboratory and hadronic c.m. systems), 
are shown in Fig.~3, and compared
with the RAPGAP simulation. Data show that the $\Dstarp$ production is 
suppressed in the central region and occurs mainly in the direction of the
virtual photon, in contrast to the standard fragmentation expectation. 
\smallskip 
\par\noindent
\begin{minipage}[t]{7cm}
\begin{minipage}[t]{6.9cm}
\epsfxsize=3.5cm\epsfbox{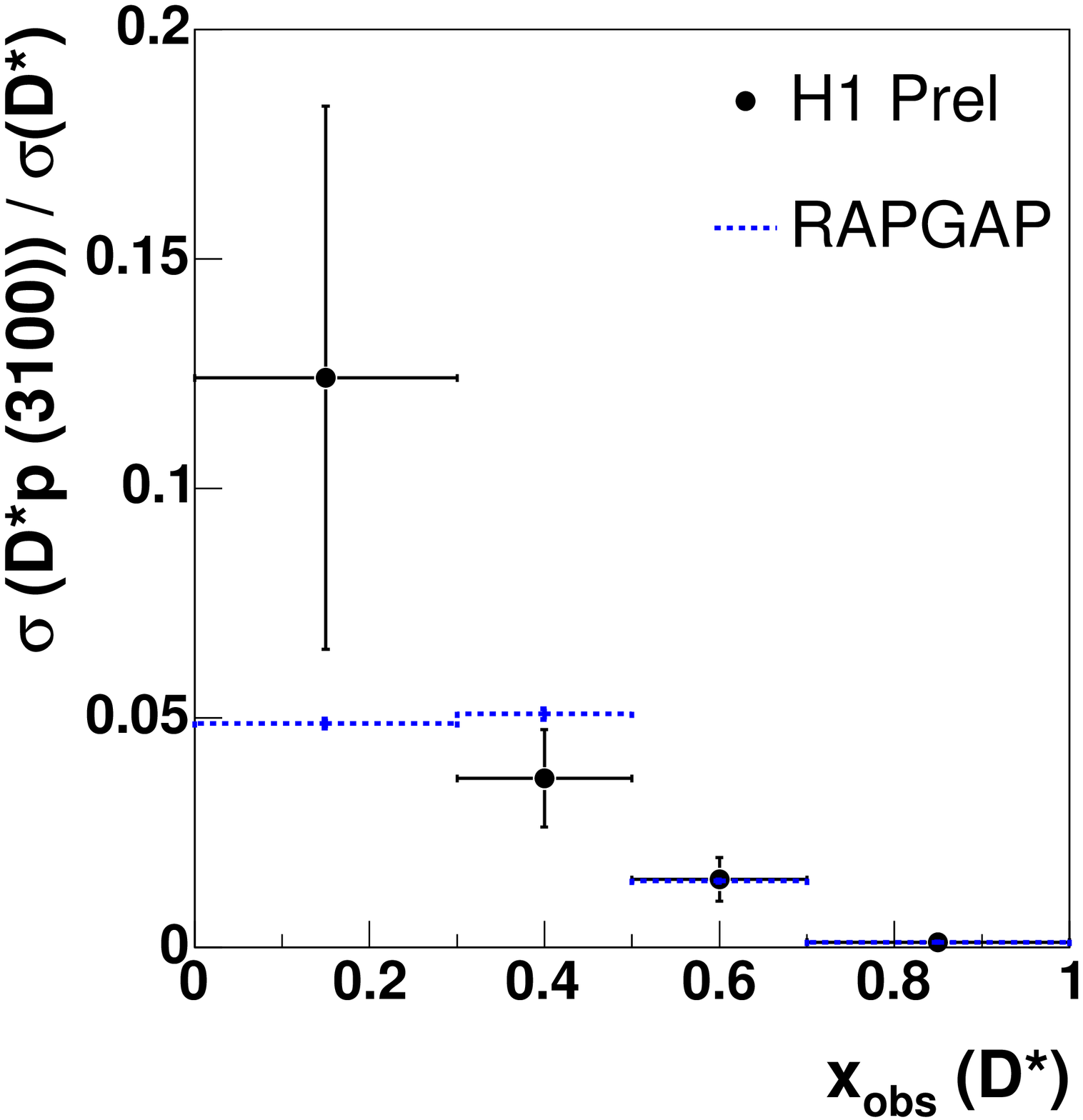}
\vspace*{-3.55cm}
\par\noindent
\hspace*{3.4cm}
\epsfxsize=3.5cm\epsfbox{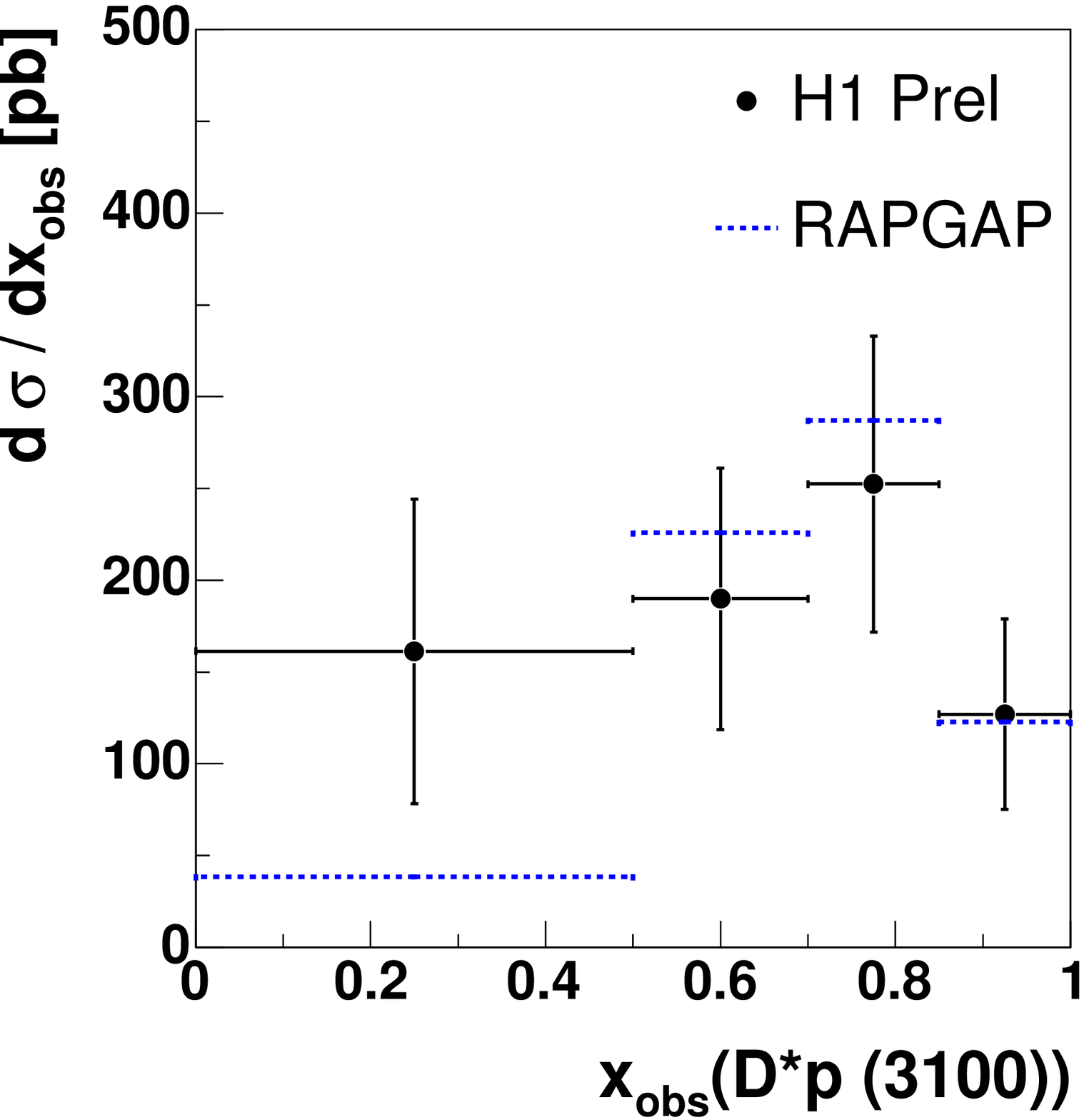}
\end{minipage}
\vspace*{-3.3cm}
\par\noindent
\hspace*{1.2cm} {\bf a)} \hspace*{3cm} {\bf b)}
\vspace*{2.9cm}
\par\noindent
\hspace*{0.4cm}
\begin{minipage}[t]{6cm}
 {Fig.~4. \  
Acceptance~corrected~ratio 
$\sigma_{vis}(\Dstarp)/\sigma_{vis}(\dstar)$ as a function of the 
$\dstar$ hadronization fraction $x_{obs}(\dstar)$ in a) and differential
cross section $d\sigma(\Dstarp)/d x_{obs}(\Dstarp)$ in b). Errors are
statistical only.
}
\end{minipage}
\end{minipage}
\smallskip
\par
In order to gain information about the $\Dstarp$ fragmentation function, 
the variable
$$
x_{obs}(charm) = \frac{(E-p_z)_{charm}}
               {\sum_{h\in hemi}(E-p_z)_h}
$$
is defined. The dependence of the
cross section ratio $\sigma(\Dstarp)/\sigma(\dstar)$ on $x_{obs}(\dstar)$ 
and the differential cross section 
$d\sigma(\Dstarp)/d x_{obs}(\Dstarp)$  are shown in  Fig.~4a and 4b, 
respectively.
The 
comparison with the RAPGAP expectation (in which $\dstar$ and 
$\Dstarp$  have the
same production mechanism) shows that the  
$\dstar$ from the $\Dstarp$ decay is softer than the inclusive $\dstar$, 
and that the 
fragmentation of $\Dstarp$ is harder than the inclusive $\dstar$ fragmentation.

\balance

\end{document}